\documentclass[aps,twocolumn,showpacs,amsmath,amssymb,superscriptaddress,longbibliography,prb]{revtex4-2}
\usepackage{physics}
\usepackage{graphicx}
\usepackage{epsf}
\usepackage{graphicx}
\usepackage{dcolumn}
\usepackage{amsmath}
\usepackage{cases}
\usepackage{color}
\usepackage[colorlinks, linkcolor=blue, citecolor=red]{hyperref}
\usepackage{soul}
\usepackage[capitalise]{cleveref}
\usepackage[dvipsnames]{xcolor}
\makeatletter
\let\old@makecaption=\@makecaption
\usepackage{subcaption}
\let\@makecaption=\old@makecaption
\makeatother
\usepackage{blindtext}
\usepackage[normalem]{ulem}
\usepackage{siunitx}

\renewcommand{\b}[1]{{\boldsymbol{#1}}}

\newcommand{\nn}{\nonumber}

\newcommand{\overbar}[1]{\mkern 1.5mu\overline{\mkern-1.5mu#1\mkern-1.5mu}\mkern 1.5mu}

\usepackage{cases}

\makeatletter
\@booleanfalse\titlepage@sw
\makeatother

\begin{document}

\title{Laguerre-Gaussian light induction of orbital currents and Kapitza stabilization in superconducting circuits }
\author{Hennadii Yerzhakov}
\affiliation{Nordita, Stockholm University, and KTH Royal Institute of Technology, Hannes Alfvéns väg 12, SE-106 91 Stockholm, Sweden}
\author{Tien-Tien Yeh}
\affiliation{Nordita, Stockholm University, and KTH Royal Institute of Technology, Hannes Alfvéns väg 12, SE-106 91 Stockholm, Sweden}
\author{Alexander Balatsky}
\affiliation{Nordita, Stockholm University, and KTH Royal Institute of Technology, Hannes Alfvéns väg 12, SE-106 91 Stockholm, Sweden}
\affiliation{Department of Physics, University of Connecticut, Storrs, Connecticut 06269, USA}

\date\today

\begin{abstract}

 We investigate the effects of a Laguerre-Gaussian (LG) beam on the superconducting state. We show that the vortex angular momentum of a LG beam  affects the superconducting state and induces currents.  The induction of the current by light is illustrated on a Josephson loop and SQUID devices. 
 In particular, we establish that coupling a dc SQUID to the AC magnetic flux of a LG beam can stabilize $\pi$ phase in the SQUID. This can happen via developing a global or local minimum in the effective potential at $\pi$. In the latter case, this happens via the Kapitza mechanism.
\end{abstract}

\maketitle

\section{Introduction}
\label{sec: Intro}

The process of interaction of quantum fields where one transfers quantum numbers from one to another is a well known phenomenon in quantum mechanics. We focus on the specific case of transfer of quantum numbers of light to coherent electron quantum state.  Examples of these transfer processes include energy, momentum, angular momentum transfers as illustrated, e.g., by light-induced momentum change of a particle and induction of magnetization due to circular polarization of the photons~\cite{pitaevskii1960,pershan1966,popova2011,popova2012,kirilyuk2010,hertel2006,majedi2021,mironov2021}
. Recently, it has been suggested that interaction of light carrying non-zero angular momentum can induce vortices in superconductors~\cite{yokoyama, sharma2023light, croitoru2022}

In this paper, we look at a newer example of the interaction of quantum matter with structured light and microwave radiation: the induction of the circular orbital  motion of the superconducting electrons illuminated by the so-called Laguerre-Gaussian (LG) beam, see~\cref{eq: LG beam electric field}. A LG beam has the vortex in light field phase~\cite{rosen2022review}. The assumption  hence is that even in linear response the phase coherent superconducting (SC) electrons will develop a coherent response while interacting  with the beam that carries vorticity. We indeed  find that LG beams do induce SC currents.  One can say that vorticity of the beam is {\em imprinted} on the coherent quantum fluid, in our case chosen to be SC fluid.   Qualitative explanation of the effect is direct: to carry orbital momentum, an electromagnetic (EM) wave propagating in $z$-direction must have its amplitude of the electric and magnetic fields to be dependent on $(x,y)$ coordinates, or in polar angle $\theta$ in case of LG beam, on which we concentrate. This means that $z$-component of the magnetic field of the beam is non-zero. This results in non-zero flux through an area normal to the beam propagation,~\cref{fig: rf}, which oscillates with the frequency of the LG beam.

The effect  is expected to be present  in broader contexts;  here we choose to illustrate it on the SQUID - the SC device that contain loop currents that are interacting with the light. 

To test these ideas  one  can use sensing of the induced flux by SQUID devices via Aharonov-Bohm like setup. We show that this can lead to the dynamically stabilized effective $\pi$ Josephson junction. While the effect we discuss has similarities with the Aharonov-Bohm effect, there is a key difference -- in the light-induced flux, the light field needs to enter into a SC and hence plays the role of the physical field that is sampled by SC current. In the traditional Aharonov-Bohm effect, the magnetic field $\b{B}$ is nonzero inside the solenoid and does not enter into the superconductor. Hence, for the setups we consider, it is important that the light field directly ``touches" the superconducting electron liquid.

The structure of the paper is as follows:  we give a brief introduction into a LG beam in~\cref{sec: LG beam flux}.  In~\cref{sec: rf SQUID}, we start  with the simpler example of a magnetic field and flux induced by a LG beam in the SC loop. In~\cref{sec: Effective JJ}, we
expand the calculation to the dc SQUID geometry. In~\cref{sec: dc SQUID}, we consider the full nonlinear dynamics of the coupled light-SQUID setup and find that, with the proper tuning, one can induce $\pi$ Josephson phase shift as a stable point of the junction potential. We also  elaborate on the light-induced Kapitza engineered $\pi$ junction. We conclude in~\cref{sec: Conclusion}.

\section{Magnetic flux generated by LG beams}
\label{sec: LG beam flux}

To the zeroth order in the paraxial parameter, the electric field of a linearly polarized LG higher mode with a wavevector $\b{k}$ and wavelength $\lambda$ propagating in $z$-direction can be considered purely transverse and written as~\cite{siegman86}

\begin{align}
\label{eq: LG beam electric field}
    \b{E}(\b{r}) = E_0(r,z) e^{i \alpha(r,z)} e^{i(k z - \omega t)} e^{i \ell \theta} \hat{\b{x}},
\end{align}
where $\b{r}=(r,\theta,z)$ is the radius-vector written in cylindrical coordinates, $\hat{\b{x}}$ is a unit vector in $x$-direction, and
\begin{align}
\label{eq: LG beam electric field explanation}
E_0(r, z) = &E_0 \frac{w_0}{w(z)}\left(\frac{r \sqrt{2}}{w(z)}\right)^{|l|} e^{-r^2/w^2(z)} L_p^{|l|}\left(\frac{2 r^2}{w^2(z)}\right), \\
\alpha(r,z) = & - k \frac{r^2}{2 R(z)} + \psi(z),
\end{align}
where $w(z)=w_0 \sqrt{1+\left(\frac{z}{z_{\mathrm{R}}}\right)^2}$ is the spot size at $z$, where $z_{\mathrm{R}}=\frac{\pi w_0^2 n}{\lambda}$ is the Rayleigh range with $n$ being the refraction index of the medium, $R(z)=z\left[1+\left(\frac{z_{\mathrm{R}}}{z}\right)^2\right]$ is the radius of curvature of the wavefront of the Gaussian beam at $z$, $\psi(z)=\arctan \left(\frac{z}{z_{\mathrm{R}}}\right)$ is the Gouy phase, and $\ell$ is an integer corresponding to the order of the LG mode. In fact, $z$-component of the electric field is not zero: From $\nabla \cdot \b{E} = 0$, it can be estimated that $E_z(r,z) \sim \frac{\lambda}{w(z)} E_x$, which is small compared to $E_x$ in the paraxial approximation.

Correspondingly, the vector potential is the real part of

\begin{align}
    \b{A}(\b{r}) = \frac{E_0(r,z)}{i \omega} e^{i \alpha(r,z)} e^{i(k z - \omega t)} e^{i \ell \theta} \hat{\b{x}},
\end{align}
and the flux through a disk-area of radius $r_0$ normal to the beam propagation is 

\begin{align}
\label{eq: LG_flux}
    \Phi_e = \oint \b{A} \cdot d{\b{l}} = \frac{E_0(r_0,z)}{\omega} e^{i \alpha(r_0,z)} e^{i(k z - \omega t)} \pi  r_0 \delta_{\ell,\pm1}. 
\end{align}
This simple observation implies the possibility of generating high-frequency oscillating electromotive force with a LG beam via Faraday effect. For modern lasers, such flux is rather small -  on the order $10^{-13}-10^{-12}$ Wb, which is only $10-10^2$ times larger than the magnetic flux quantum, $\Phi_0$. However, the Josephson phase is sensitive to the magnetic flux which is on the order of $\Phi_0$, and in the next sections we study the interaction of such a flux from LG beams with superconducting rings with Josephson junctions (JJ).

We note that the flux is contour dependent, and specifically chosen contours for other type of beams also can contain non-zero flux. However, a circular-shaped contours look particularly convenient to us.

\section{A LG beam shone on a rf SQUID}
\label{sec: rf SQUID}

\begin{figure}[!htbp]
    \includegraphics[width=1\columnwidth]{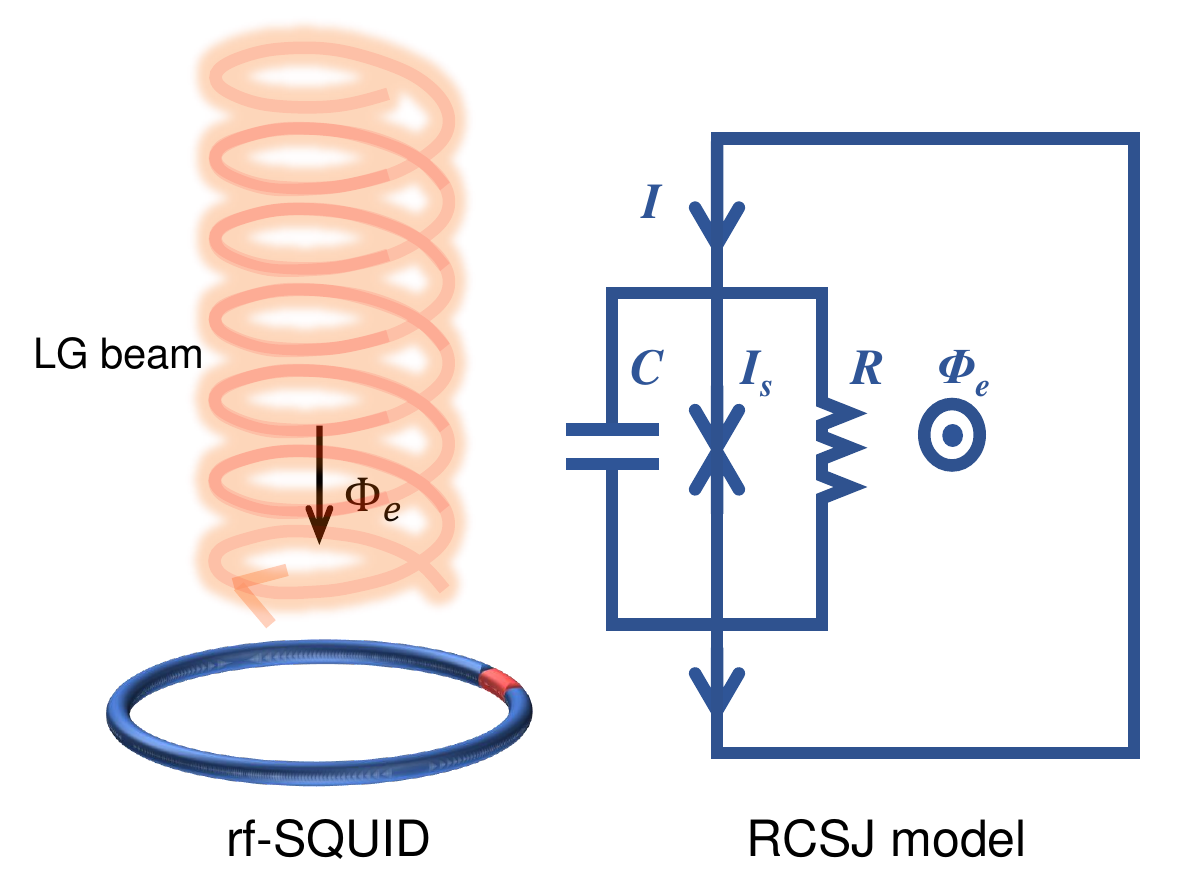}
    \caption{A LG beam shone into an rf SQUID (left) and the equivalent scheme (right).}
    \label{fig: rf}
\end{figure}

We start with shining LG light into the orifice of an rf SQUID structure, see~\cref{fig: rf}. In the following, we neglect the physics that can lead to Shapiro effect~\cite{shapiro1963} due to the electric component of the EM wave and photon-assisted tunneling~\cite{dayem1962, tien1963}. This is a reasonable approximation if the JJs are positioned in such a way that the polarization of the electric field is perpendicular to the Josephson current through the junctions (i.e., it is parallel to the surfaces of the electrodes that constitute the junctions) and the power of the beam is sufficiently small.

The RCSJ model for the dynamics of the rf SQUID then reads~\cite{likharev2022dynamics}
\begin{align}
\label{eq: RCSJ_rf}
    I &=C \frac{d^2 \Phi}{d t^2}+\frac{1}{R} \frac{d \Phi}{d t}+I_0 \sin \left(2 \pi \frac{\Phi}{\Phi_0}\right), \\ \nn
    \Phi &= \Phi_e - LI,
\end{align}
where $\Phi$ is the full flux through the orifice of the SQUID, $C$, $R$, and $L$ are the capacitance, resistance, and self-inductance of the rf SQUID, respectively, $I_0$ is the Josephson critical current, $I$ is the full current in the loop, and $\Phi_0$ is the magnetic flux quantum. Combined, eqs. in~\cref{eq: RCSJ_rf} give
\begin{align}
\label{eq: rf-SQUID}
    \ddot \phi + \Gamma \dot \phi + \frac{\omega_0^2}{2\pi} \sin (2 \pi \phi)  = \omega_L^2 \phi_e - \omega_L^2 \phi,
\end{align}
where $\phi = \Phi/\Phi_0$, $\phi_e = \Phi_e/\Phi_0$, $\omega_0^2= 2 \pi I_0/\Phi_0 C$, $\omega_L^2 = 1/LC$, and $\Gamma = 1/RC$. This model with oscillating in time $\phi_e$ studied in~\cite{Hizanidis2016,Hizanidis2018,Hizanidis2020} exhibits complex dynamic behavior including chaos.

\section{A LG beam is shone inside a dc SQUID shunted by an external capacitor}
\label{sec: Effective JJ}

\begin{figure}[!htbp]
    \includegraphics[width=0.99\linewidth]{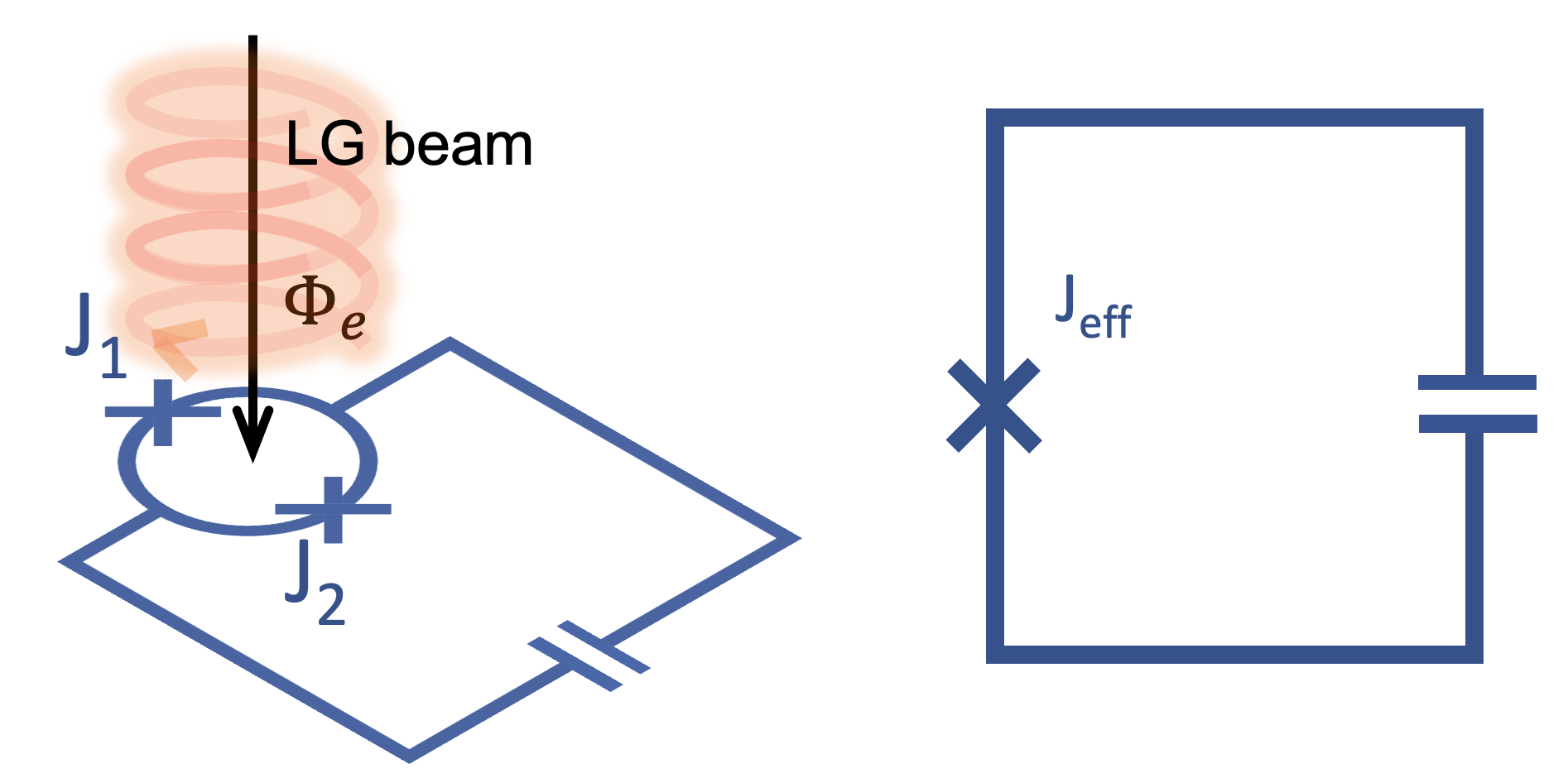}
    \caption{A LG beam shone into a dc SQUID shunted by an external capacitor (left)  and the equivalent scheme (right).}
    \label{fig: Scheme}
\end{figure}

We can couple $\phi_e(t)$ to the dynamic variable $\phi$ in the dc SQUID setup.
Let us consider a symmetric dc SQUID shunted by an external capacitor as depicted in~\cref{fig: Scheme}. We start with the simplest model neglecting the self-inductances of the electrodes comprising the dc SQUID and the capacitances and resistances of the JJs. In this case, the dc SQUID behaves effectively like a JJ with the amplitude of the Josephson critical current modulated by the external flux through the dc SQUID~\cite{Hatridge2011}.

Indeed, the flux through the SQUID is
\begin{align}
\label{eq: flux_in_dc-SQUID}
    \phi_1-\phi_2=2 \pi \phi_e - \frac{2 \pi}{\Phi_0} 2L I_c,
\end{align}
where $\phi_{1,2}$ are phase differences across the JJs, $\phi_e=\frac{\Phi_e}{\Phi_0}$, and $I_c$ is the current circulating in the loop of the dc SQUID. Neglecting $2LI_c$ term in~\cref{eq: flux_in_dc-SQUID}, the interference of the currents $I_1$ and $I_2$ through JJs gives
\begin{align}
\label{eq: Iinterference}
    I=I_1+I_2=2 I_0 \cos (\pi \phi_e) \sin \phi_+,
\end{align}
where $\phi_+=\frac{\phi_1+\phi_2}{2}$.

Ultimately, we can think about the dc SQUID as an effective JJ with flux modulated critical current $I_{0,eff}(\phi_e)=2 I_0 \cos (\pi \phi_e)$. Under the assumptions, the voltage across the SQUID is
\begin{align}
\label{eq: Vsquid}
    V=\frac{\Phi_0}{2\pi} \dot \phi_+.
\end{align}
Combinination of~\cref{eq: Iinterference,eq: Vsquid}, and $I+\dot Q =I_b$, where $Q$ is the charge on the shunting capacitor and $I_b$ is the bias current (which we do not show in~\cref{fig: Scheme}) assumed to be constant, leads to the equation for $\phi_+$ analogous to the RCSJ model
\begin{align}
\label{eq: Oscillation_eq0}
    \ddot \phi_+ + \Gamma \dot \phi_+ + \omega^2(\phi_e) \sin \phi_+ = \omega_0^2 \frac{I_b}{2 I_0},
\end{align}
where we have also added a dissipative term $\Gamma \dot \phi$ and $\omega^2(\phi_e) = \frac{2 \pi I_{0,eff}(\phi_e)}{C \Phi_0} = 2 \omega_0^2 \cos (\pi \phi_e)$, where $\omega_0^2=\frac{2 \pi I_0}{C \Phi_0}$. 

We assume now, according to~\cref{eq: LG_flux}, the time-dependence for the external flux is cosinusoidal, such that $\phi_e(t) = \phi_0 \cos(\omega t)$ with $\phi_0 = \max{\Phi_e/\Phi_0}$, and $I_b=0$. 
Using the Jacobi–Anger expansion~\cite{abramowitz+stegun} 
\begin{align}
    \cos (x \cos\alpha) = J_0(x) + 2 \sum_{p=1}^\infty (-1)^p J_{2p}(x) \cos(2 p \alpha),
\end{align}
we rewrite~\cref{eq: Oscillation_eq0} as
\begin{align}
\label{eq: Oscillation_eq1}
    \ddot \phi + \Gamma \dot \phi + 2 \omega_0^2 J_0(\phi_0) \sin \phi &= \\ \nn
    -4 \omega_0^2 \sum_{p=1}^\infty (-1)^p J_{2p}(\phi_0) \cos(2 p \omega t) \sin \phi,
\end{align}
which we recognize as an equation of an oscillator in a fast oscillating field. Then, for sufficiently large $\omega$, we can expect the Kapitza-pendulum scenario~\cite{kapitzaor1,*kapitza1,kapitzaor2,* kapitza2, landau1982mechanics}, i.e., the stabilization upon drive  of the formerly unstable state. We investigate such a possibility in the next section, where, based on the simple idea layed out in this section, we systematically study Kapitza-pendulum scenario in the dc SQUID geometry without shunting the SQUID by an external capacitor. The case with an external capacitor without neglecting the JJ's capacitances and inductances is briefly discussed in~\cref{app: dc-SQUID_shunted}.

\section{dc SQUID}
\label{sec: dc SQUID}

Based on the ideas in the previous section, we investigate the possibility of the Kapitza scenario in a dc SQUID subjected to the flux from a LG beam without it being shunted by an external capacitor. A vast literature is devoted to the study of dynamics in dc SQUIDs~\cite{wang2023, tuckerman1980, zappe1978, faris1981, ben-jacob1981, ryhanen1992, peterson1983, polak2007}. Kapitza engineering~\cite{voronova2016,martin2018,lerose2019,kulikov2022,kuzmanovski2022} of a SQUID has not been widely discussed to date. 

\begin{figure}[!htbp]
    \includegraphics[width=0.8\columnwidth]{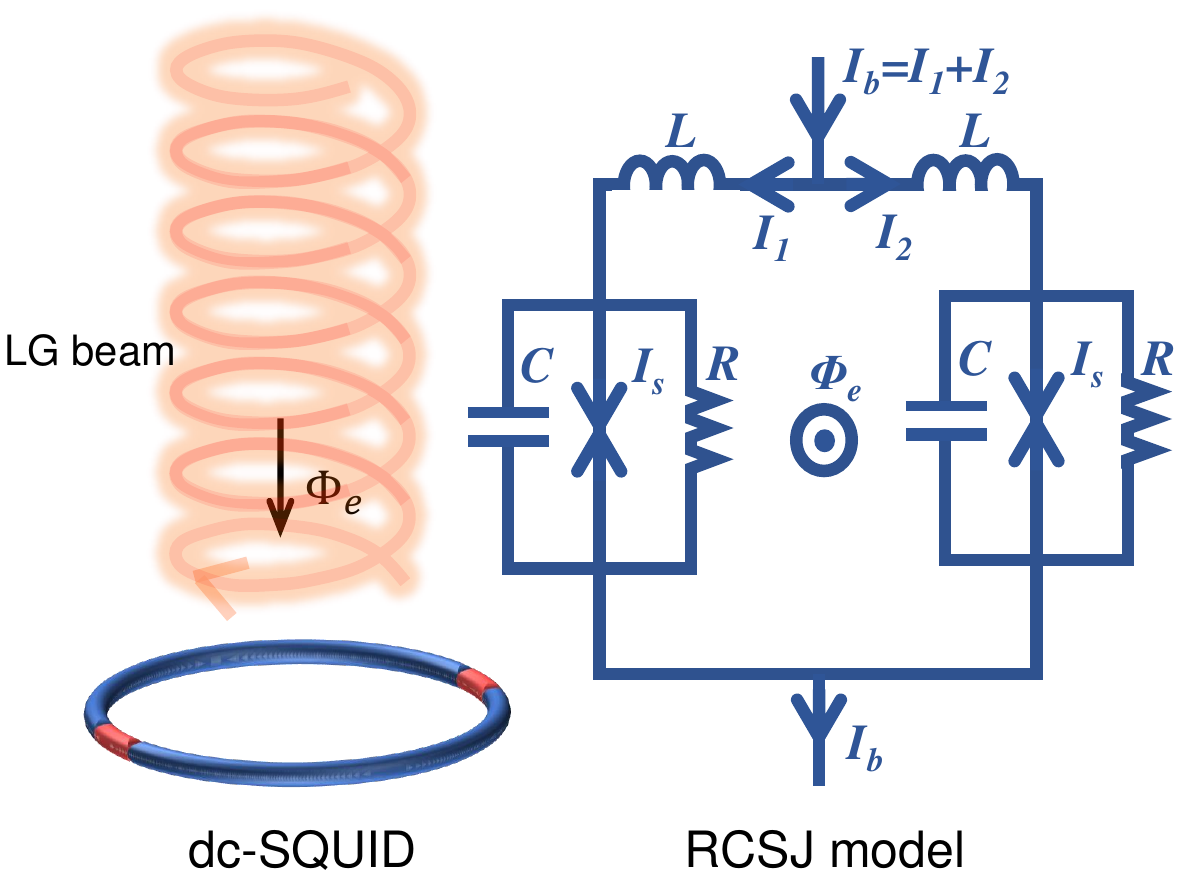}
    \caption{A LG beam shone into a dc SQUID (left)  and the equivalent scheme extended to account for the bias current $I_b$ (right).}
    \label{fig: dc-SQUID}
\end{figure}

Consider an equivalent scheme of a symmetric dc SQUID  depicted in~\cref{fig: dc-SQUID}. The relation for the flux and phases reads

\begin{align}
\label{eq: flux_relation}
    \frac{\Phi_0}{2 \pi} \left( \phi_1-\phi_2 \right)=\Phi_e - 2 L I_c,
\end{align}
where $I_c=(I_1-I_2)/2$ is the current circulating in the loop of the dc SQUID. For the time-dependent $\phi_e$, this equation assumes instantaneous adjustment of the phases $\phi_{1,2}$. This is approximately true if the time-scale on which $\phi_e$ changes is significantly slower than the relevant time-scale for the superconductor dynamics, $\tau_r \sim \hbar/\Delta$, where $\hbar$ and $\Delta$ are the Plank constant and the superconducting order parameter, respectively.

The set of Kirhoff's laws for currents reads

\begin{align}
\label{eq: Kirhoff_dc-SQUID}
    I_0 \sin \phi_1 + \frac{\Phi_0}{2 \pi R} \dot \phi_1 + C \frac{\Phi_0}{2 \pi} \ddot \phi_1 = I_1, \\ \nn
    I_0 \sin \phi_2 + \frac{\Phi_0}{2 \pi R} \dot \phi_2 + C \frac{\Phi_0}{2 \pi} \ddot \phi_2 = I_2, \\ \nn
    I_1+I_2 = I_b,
\end{align}
where $I_b$ is the bias current assumed to be constant.
And the voltage across the dc SQUID is given by (see~\cref{app: dc-SQUID_shunted} for an example of the derivation)

\begin{align}
\label{eq: dc-SQUID_voltage}
    V=\frac{\Phi_0}{2 \pi} \frac{\dot \phi_1+ \dot \phi_2}{2} - \frac{L}{2} \dot I_b. 
\end{align}

Taking sum and difference of the first two equations in~\cref{eq: Kirhoff_dc-SQUID} and introducing notation $\phi_{\pm} = \frac{\phi_1 \pm \phi_2}{2}$, $\omega_0^2=\frac{2 \pi I_0}{\Phi_0 C}$, $\Gamma = \frac{1}{CR}$, and $\omega_L^2 = \frac{1}{LC}$, we obtain

\begin{align}
\label{eq: phi_plus_minus}
     & \ddot \phi_+ + \Gamma \dot \phi_+ + \omega_0^2 \cos(\pi \phi) \sin \phi_+ = \omega_0^2 \frac{I_b}{2 I_0}, \\ \nn
     & \ddot \phi + \Gamma \dot \phi + \frac{\omega_0^2}{\pi} \cos\phi_+ \sin (\pi \phi)  = \omega_L^2 \phi_e - \omega_L^2 \phi, 
\end{align}
where $\phi = \frac{\Phi}{\Phi_0}$ is the full flux in units of $\Phi_0$ and relates to $\phi_-$ via~\cref{eq: flux_relation}.

Note that the second equation in~\cref{eq: phi_plus_minus} is an equation for the flux inside the SQUID and is essentially the same as~\cref{eq: rf-SQUID} for the dynamics of the flux inside an rf SQUID.

We can deem the considered system as a black-box device and ask a question about its I-V characteristic. Under typical assumption that $I_b$ is created by an external source of constant current, the output voltage, according to~\cref{eq: dc-SQUID_voltage}, is $V=\frac{\Phi_0}{2 \pi} \dot \phi_+$. Thus, in the Kapitza scenario of inverted pendulum, where $\phi_+$ oscillate around $\pi$, such a black-box device may be considered as an effective $\pi$-phase Josephson junction.

For $L=0$, two equations in~\cref{eq: phi_plus_minus} decouple. We start with this case in the following section.

\subsection{Neglecting the self-inductance: $L=0$}

When $L=0$, $\phi$ is simply given by the external flux: $\phi = \phi_e$. For generality, we will include a dc component of the external flux

\begin{align}
\label{eq: phi_external+dc}
    \phi_e=\phi_1 + \phi_0 \cos(\omega t).
\end{align}
Substituting this into the first equation in~\cref{eq: phi_plus_minus} and using the Jacobi-Anger expansion, we obtain

\begin{align}
\label{eq: dc-SQUID_L=0}
    \ddot \phi_+ &+ \Gamma \dot \phi_+ + \omega_0^2 \cos \pi \phi_1 J_0(\pi \phi_0) \sin \phi_+ = \\ \nn
    &-2 \omega_0^2 \sum_{p=1}^\infty J_{p}(\pi \phi_0) \cos(\pi \phi_1+\frac{\pi}{2} p) \cos(p \omega t) \sin \phi_+,
\end{align}
where we also set $I_b=0$. Non-zero $I_b$ cases then can be taken into account in the context of the washboard potential.

To investigate this, we employ the idea of separation of time scales~\cite{kapitzaor1,*kapitza1,kapitzaor2,*kapitza2,landau1982mechanics}: given $\frac{\omega}{\omega_0 \sqrt{\abs{J_0(\phi_0)}}} \gg 1$, we expect the  solution $\phi_+(t)$ can be described as a slowly varying component (on the scale of $\left(\omega_0 \sqrt{\abs{J(\phi_0)}} \right)^{-1}$) with small-amplitude fast oscillations (on the scale of $\omega^{-1}$) on top of it~\footnote{An analytical method of wider applicability can be used~\cite{butikov2011}}, i.e., we represent
\begin{align}
    \phi_+=\theta + \psi,
\end{align}
where $\theta$ and $\psi$ are the slow and fast components, respectively. Substituting this in~\cref{eq: dc-SQUID_L=0}, expanding in $\psi$, and collecting the dominant fast terms together, we obtain an equation for $\psi$

\begin{multline}
\label{eq: Oscillation_eq_fast}
    \ddot \psi + \Gamma \dot \psi =\\ -2 \omega_0^2 \sum_{p=1}^\infty J_{p}(\pi \phi_0) \cos(\pi \phi_1+\frac{\pi}{2} p) \cos(p \omega t) \sin \theta,
\end{multline}
where we kept the dissipative term in~\cref{eq: Oscillation_eq_fast} to incorporate the overdamped regime, $\Gamma \gg \omega_0 \sqrt{\abs{J(\phi_0)}}$. The solution for the fast component is 

\begin{align}
    \psi(t) = &\sum_{p=1}^\infty \frac{2 \omega_0^2 J_{p}(\pi \phi_0) \cos(\pi \phi_1 + \frac{\pi}{2}p)}{\sqrt{p^4 \omega^4 +\Gamma^2 p^2 \omega^2}} \cos(p \omega t +\alpha_p) \sin \theta \nn \\ 
    &+ A e^{-\Gamma t} +B,
\end{align}
where the phase shifts $\alpha_p$ satisfy $\tan \alpha_p = \Gamma/(p \omega)$. We set $A=B=0$ to comply with the physical picture of the solution (small-amplitude fast oscillations about the slow-mode component).

For the slow term, after averaging over time, we have

\begin{align}
\label{eq: Oscillation_eq_slow}
    \ddot \theta + \Gamma \dot \theta + \omega_0^2 \cos \pi \phi_1 J_0(\pi \phi_0) \sin \theta = \\ \nn 
    -2\sum_{p=1}^\infty \frac{\omega_0^4 J_{p}^2(\pi \phi_0) \cos^2(\pi \phi_1 + \frac{\pi}{2}p)}{\sqrt{p^4 \omega^4 +\Gamma^2 p^2 \omega^2}} \sin \theta \cos \theta,   
\end{align}

When $\cos \phi_1 J_0(\pi \phi_0)>0$,~\cref{eq: Oscillation_eq_slow} is an equation of motion for a mathematical pendulum in an effective potential

\begin{align}
\label{eq: U_eff}
    U_{eff}=\omega_0^2 \cos \pi \phi_1 J_0(\pi \phi_0) \left( -\cos \theta + K \sin^2 \theta \right),
\end{align}
where
\begin{align}
\label{eq: Kapitza_factor}
    K=\frac{\omega_0^2}{\omega^2} \sum_{p=1}^\infty \frac{J_{p}^2(\pi \phi_0) \cos^2(\pi \phi_1 + \frac{\pi}{2}p)}{p^2\sqrt{1 +\Gamma^2/ (p^2 \omega^2)} J_0(\pi \phi_0) \cos \pi \phi_1}.
\end{align}

This potential develops a local minimum at $\theta = \pi$ when $K>\frac{1}{2}$. It is easy to see that such a condition can be satisfied for $\pi \phi_0$ sufficiently close to the roots of the zero-order Bessel function of the first kind or/and for $\pi \phi_1$ such that $\cos \pi \phi_1$ is sufficiently small. Simultaneously, the natural frequency of the oscillator in~\cref{eq: Oscillation_eq_slow} becomes small, which allows for the Kapitza effect to occur at frequencies $\omega$ comparable to $\omega_0$, the plasma frequency of the Josephson junction.

\begin{figure*}[tbp]
    \includegraphics[width=0.99\textwidth]{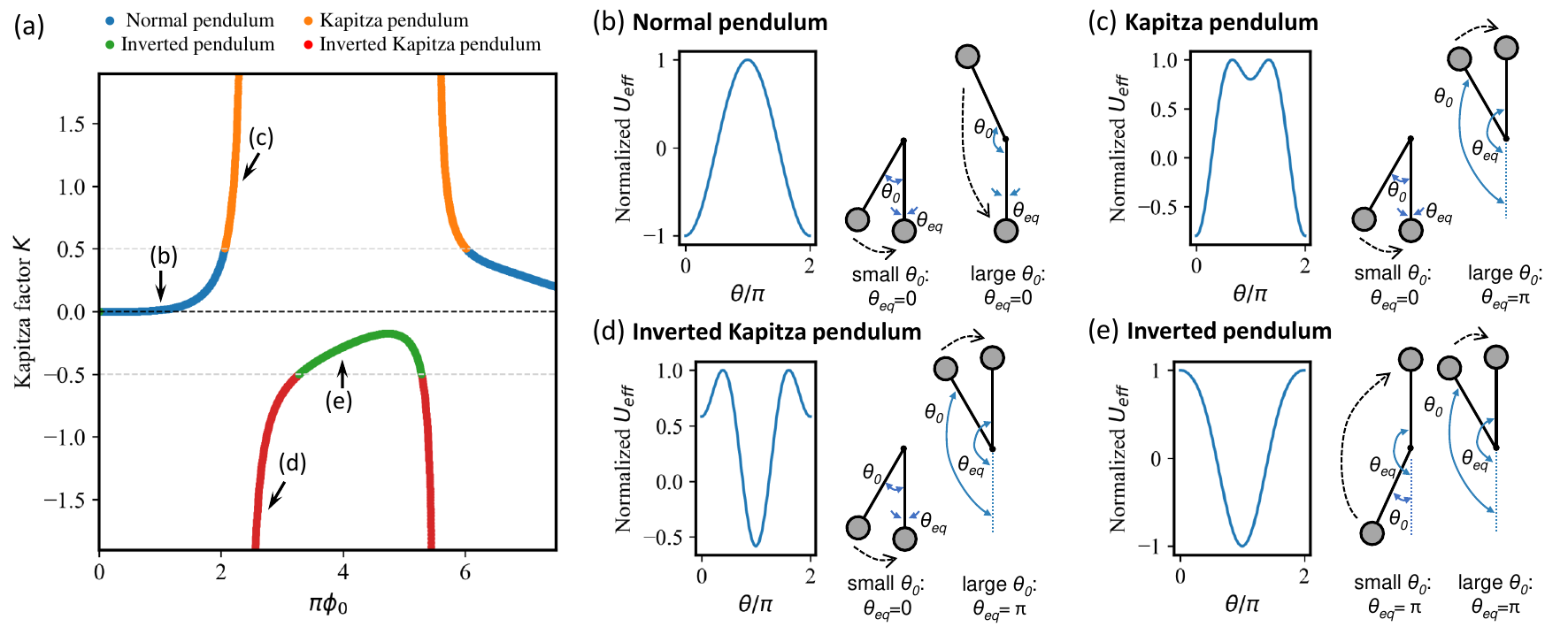}
    \caption{a) Kapitza factor, $K$, dependence on $\pi \phi_0$ at $\omega=\omega_0$; b), c), d), e) Schematic illustrations of different pendulum regimes with effective potential $U_{eff}$.}
    \label{fig: Kapitza_regimes}
\end{figure*}

When $\cos \pi \phi_1 J_0(\pi \phi_0)<0$, the coefficient in the third term in~\cref{eq: Oscillation_eq_slow} becomes negative. This sign can be absorbed by shifting $\theta$ by $\pi$ in~\cref{eq: Oscillation_eq_slow}: $\theta \rightarrow \theta + \pi$. Correspondingly, the effective potential shifts by $\pi$ as well, moving the global minimum to $\theta=\pi$. We call this regime the inverted pendulum regime. The Kapitza effect, i.e., the stabilization of the oscillation around the emergent local minimum, then is realized for $K<-\frac{1}{2}$. We call this regime the inverted Kapitza pendulum regime. Setting $\pi \phi_1=0$, we illustrate these regimes in~\cref{fig: Kapitza_regimes}, where we plot $K$-vs-$\pi \phi_0$ dependence at $\omega=\omega_0$.

\begin{figure}[!htbp]
    \includegraphics[width=0.99\columnwidth]{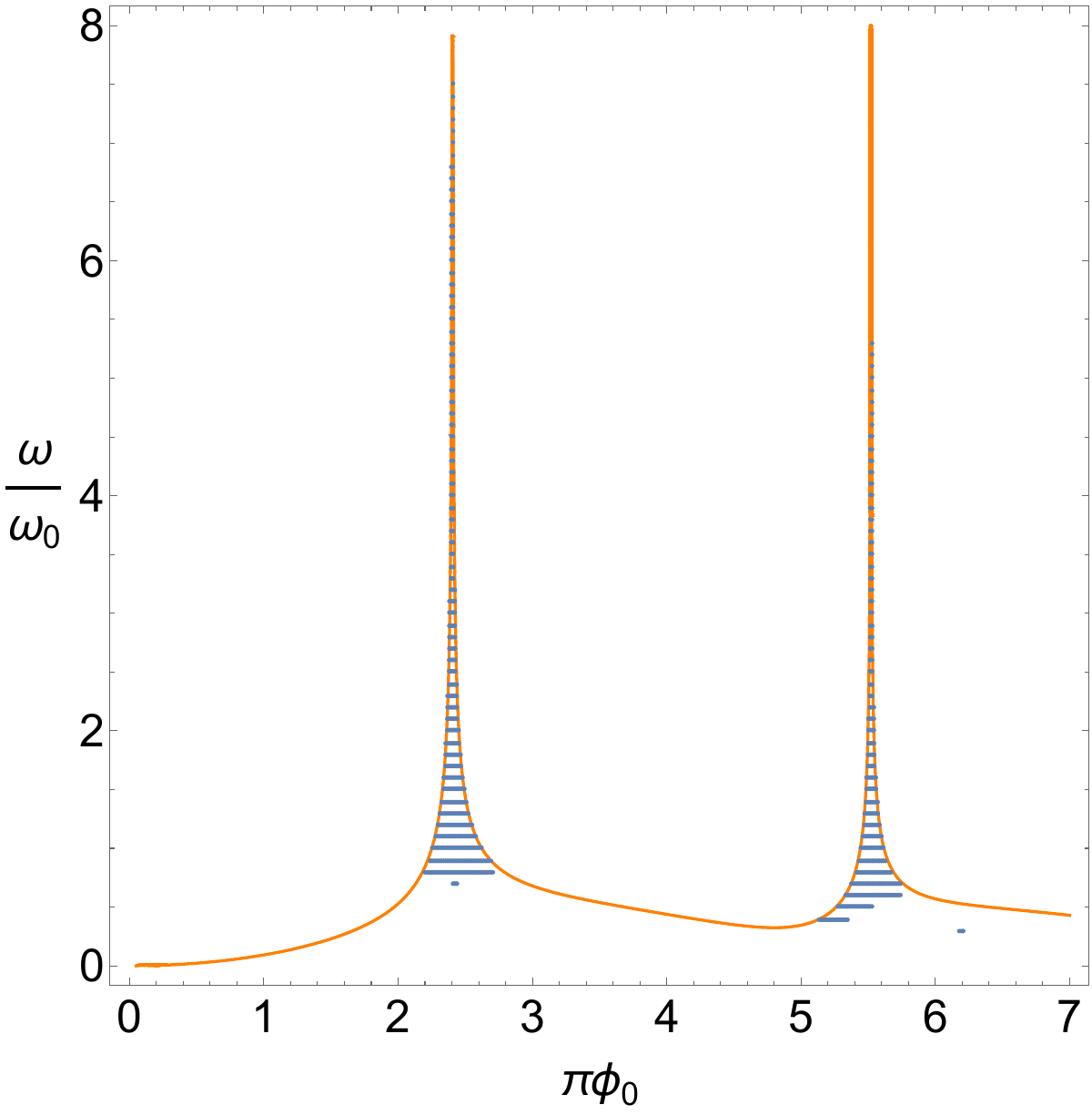}
    \caption{$K=\frac{1}{2}$ theoretical curve (orange) compared to numerically computed region (set of blue dots) in $\pi \phi_0$-$\omega/\omega_0$ plane at zero dc flux, in which the Kapitza effect takes place. For numerical calculations we set $\Gamma=0.01 \omega_0$. We also use discretization in the $\pi \phi_0$ direction with a step $0.007$ and in the $\omega/\omega_0$ direction with a step $0.1$.}
    \label{fig: KapitzaRegionNumerical}
\end{figure}

\begin{figure}[!htbp]
    \includegraphics[width=0.99\columnwidth]{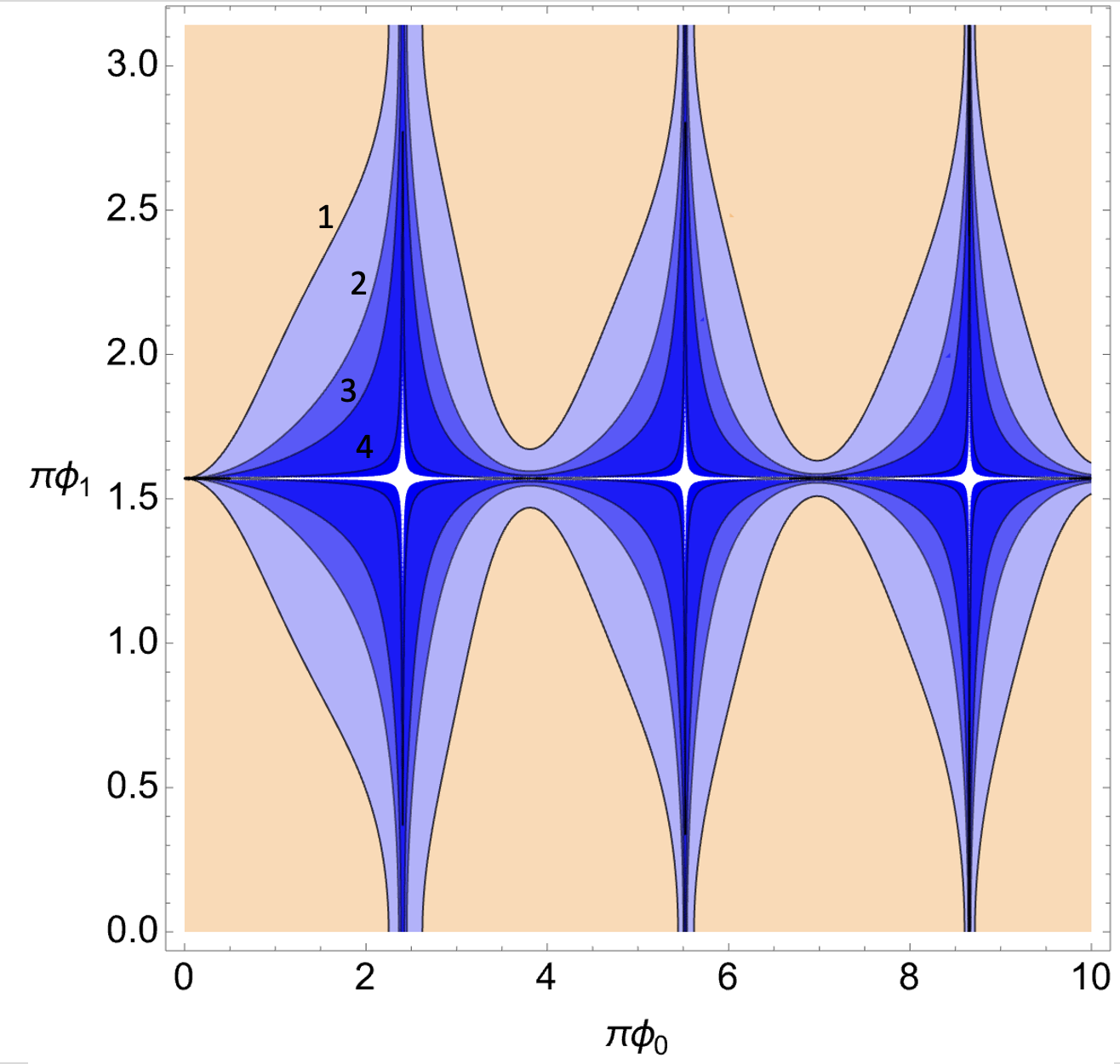}
    \caption{Theoretical regions of the Kapitza effect (blue-shaded regions and the white colored region) in $\pi \phi_0$-$\pi \phi_1$ plane. The increase of intensity of blue color corresponds to values of $\omega/\omega_0=1,2,3,10$ in increasing order. Lighter shaded regions fully contain more intense shaded areas. The labeled curves $1,2,3,4$ are curves for which $K=\frac{1}{2}$ at $\omega/\omega_0=1,2,3,10$, respectively. The white region is also the region of the Kapitza effect and indicates the divergence of $K$.}
    \label{fig: KapitzaRegion_phi0_phi1}
\end{figure}

Found analytically under the assumptions explained above,~\cref{eq: Kapitza_factor} is not exact and does not allow to determine the smallest values of $\omega$ at which the Kapitza effect still occur. In~\cref{fig: KapitzaRegionNumerical}, we plot a curve in $\pi \phi_0$-$\omega$ plane at $\pi \phi_1=0$ at which $K=\frac{1}{2}$ and numerically determined region, where the Kapitza effect takes place
. We see that the theoretical curve fit the numerically determined boundaries of Kapitza region quite well for sufficiently large $\omega$. In~\cref{fig: KapitzaRegion_phi0_phi1}, we plot $K=\frac{1}{2}$ curves in $\pi \phi_0$-$\pi \phi_1$ plane for four values of $\omega$: $\omega/\omega_0=1,2,3,10$.

\subsection{Relevant parameters of the LG beam}

The model we have considered above have certain bounds of applicability. 
As follows from the discussion after~\cref{eq: flux_relation}, the frequency of $\phi_e$ should satisfy $\omega \ll \Delta/\hbar$.

Considering typical $T_c$ in high-temperature superconductors being about $100$ K, the frequency of the LG beam then has to satisfy $\omega \ll 10^{12}$ Hz, i.e., $\omega$ has to lie in the microwave range. Correspondingly, the wavelength of the LG beam is at least on the order of $10^{-3}$ m. Consequently, the waist of the beam is also at least $10^{-3}$ m. This would likely imply that in a realistic experiment the paraxial parameter $f=\frac{\lambda}{2 \pi w_0}$ is not small, and the formulas in~\cref{eq: LG beam electric field,eq: LG beam electric field explanation} are not quantitatively correct. Nevertheless, the value of the transverse component remains of the same order as the paraxial approximation predicts~\cite{agrawal79}. Thus, since the total flux through the ring is determined by the transverse component of the vector potential, we proceed with the estimation of the flux through the ring at $z=0$ (i.e., at the beam's waist) using the electric field given by~\cref{eq: LG beam electric field}

\begin{align}
\label{eq: Maser flux}
    \pi \phi_0= \frac{E_0 w_0}{\omega \Phi_0} f(r),
\end{align}
where $f(r)=\sqrt{2} \pi^2 \left( \frac{r}{w_0} \right)^2 e^{-r^2/w_0^2}$, which has a maximum at $r/w_0$ = 1.

According to~\cref{eq: Kapitza_factor}, the smallest values of flux at which the  Kapitza effect is observed is around $\pi \phi_0 \approx 2.405$, the first root of $J_0(\pi \phi_0)=0$. And thus, from~\cref{eq: Maser flux}, we find that for $r/w_0=1$ this value is achieved at $E_0 w_0 \approx 10^{-15} [\mathrm{V}\cdot \mathrm{s}] \cdot \omega$, which for $\omega = 10^{10}$ Hz becomes $E_0 w_0 \approx 10^{-5}$ V.

In the paraxial approximation, the total power of the $\ell=1$ LG mode is given by

\begin{align}
\label{eq: Power output}
    P=\frac{\pi (E_0 w_0)^2}{4 \mu_0 c},
\end{align}
which is the same formula as for the power of the fundamental Gaussian mode.

The power output of contemporary continuous-wave masers is much lower than that of lasers, typically on the order of picoWatts. For $P \sim 10^{-12}$ W, from~\cref{eq: Power output}, we obtain $E_0 w_0 \sim 2 \cdot 10^{-5}$ V, which is enough to achieve the desirable amount of flux. In case of a maser based on nitrogen vacancies in diamond, which produces one of the highest output power of $P \approx 2.2 \cdot 10^{-9}$ W~\cite{zollitsch2023, jin2015}, we estimate
$E_0 w_0 \approx 2 \cdot 10^{-3}$ V.

\subsection{The effect of self-inductance: $L \neq 0$}

When $L \neq 0$, two equations in~\cref{eq: phi_plus_minus} are coupled. Here we discuss the results of numerical investigation of this set of equations at non-zero $L$.

\begin{figure*}[tbp]
    \includegraphics[width=0.90\textwidth]{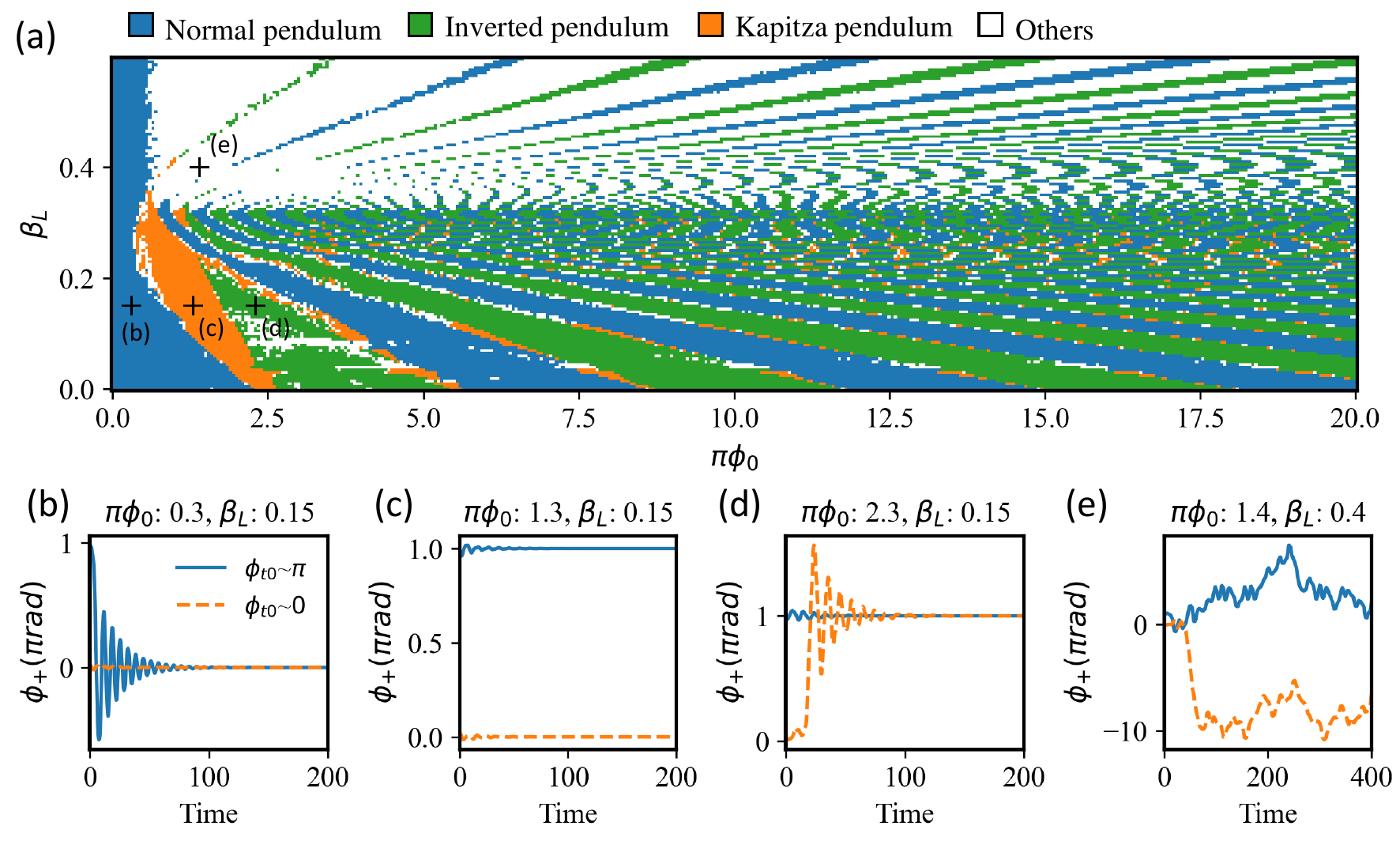}
    \caption{a) The phase diagram of the regime of oscillations for $\omega$/$\omega_{0}=1$. The white color marked ``Others" represent the behaviors not characterized as normal, Kapitza, or inverted pendulum regimes.
    b) -- e) The behavior of $\phi_{+}$ as a function of time (as only the general behavior is under interest, units of time are not indicated) for different fluxes of LG beam and screening parameter $\beta_{L}$. The corresponding points on the phase diagram are indicated by crosses. The discretization in the $\pi \phi_0$ direction is $\delta \pi \phi_0 = 0.05$, and in the $\beta_L$ direction is $\delta \beta_L = 0.005$. We set $\Gamma=0.1 \omega_0$ and $\phi_e(t) = \phi_0 \sin (\omega t)$ in computing this phase diagram. For the initial conditions, we have used $\phi_+(0) = 0.98 \pi$ and $\phi_+(0) = 0.02 \pi$ for blue and orange lines, respectively, $\dot \phi_+(0) = 0$, $\phi(0) = 0$, $\dot \phi(0) = \phi_0 \omega$.}
    \label{fig: Phase diagram1}
\end{figure*}

\begin{figure*}[tbp]
    \includegraphics[width=0.90\textwidth]{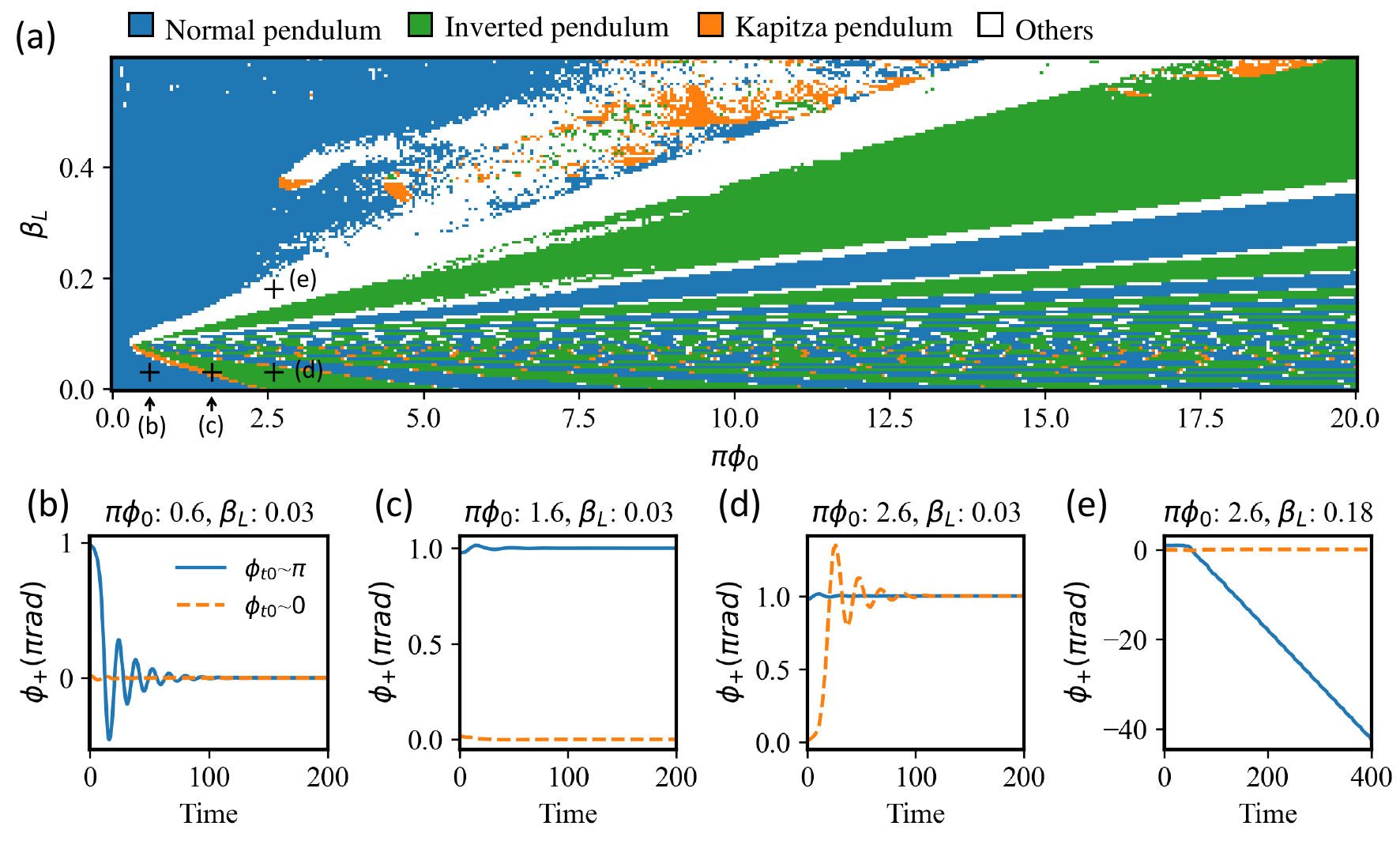}
    \caption{a) The phase diagram of the regime of oscillations for $\omega$/$\omega_{0}=2$. The white color marked ``Others" represent the behaviors not characterized as normal, Kapitza, or inverted pendulum regimes.
    b) -- e) The behavior of $\phi_{+}$ as a function of time (as only the general behavior is under interest, units of time are not indicated) for different fluxes of LG beam and screening parameter $\beta_{L}$. The corresponding points on the phase diagram are indicated by crosses. The discretization in the $\pi \phi_0$ direction is $\delta \pi \phi_0 = 0.05$, and in the $\beta_L$ direction is $\delta \beta_L = 0.005$. We set $\Gamma=0.2 \omega_0$ and $\phi_e(t) = \phi_0 \sin (\omega t)$ in computing this phase diagram. For the initial conditions, we have used $\phi_+(0) = 0.98 \pi$ and $\phi_+(0) = 0.02 \pi$ for blue and orange lines, respectively, $\dot \phi_+(0) = 0$, $\phi(0) = 0$, $\dot \phi(0) = \phi_0 \omega$.}
    \label{fig: Phase diagram2}
\end{figure*}

If $L$ is non-negligible, then the screening effect from the current inside the closed-loop SQUID circuit affects the dynamics of the effective phase $\phi_+$. We utilize the Crank–Nicolson method to solve for the time trace of the phase $\phi_{+}$ and flux $\phi$ in~\cref{eq: phi_plus_minus}. Adopting the screening parameter $\beta_{L} = \frac{2L I_0}{\Phi_0}$, we plot the phase diagram in terms of the normal, Kapitza (encompassing both the Kapitza and the inverted Kapitza), and inverted pendulum regimes in~\cref{fig: Phase diagram1,fig: Phase diagram2} for $\omega/\omega_0=1$ and $\omega/\omega_0=2$, respectively, without distinguishing between the Kapitza and the inverted Kapitza regimes.

For both ratios of $\omega/\omega_0$ after small finite value of $\pi \phi_0$, the phase diagrams display ray-like structures of normal and inverted regimes with origins on the horizontal line at finite $\beta_L$. These ``rays" are separated by the ``rays" with a mix of the Kapitza and ``Other" regimes (see~\cref{fig: Phase diagram1,fig: Phase diagram2}). Here, it is possible that some of the inclusions (especially in the lower part of the plots) of regions of ``Other" regimes is due to instabilities in our numerical solutions. For $\pi \phi_0 < \pi \phi_0^*$ ($\approx 0.36$ for $\omega/\omega_0=1$ and $\approx 0.31$ for $\omega/\omega_0=2$), there is only normal pendulum regime of the dynamics of $\phi_+$.

\section{Conclusion and discussion}
\label{sec: Conclusion}

In this study we present the results for stirring of the circular motion of the superconducting fluid with structured light, a LG beam. In particular, we find Laguerre-Gaussian beams with $\ell=1$ provide non-zero AC magnetic flux through a disk-area of finite radius. This flux can be used to induce currents in superconducting rings and superconducting rings with Josephson junctions.

In the dc SQUID geometry, a SC ring with two JJ, we have found that coupling of periodic flux might lead to the Kapitza effect -- the stabilization of oscillations about emergent local minimum at phase equal to $\pi$. In the limit of negligible self-inductance, the ranges of the flux amplitude for which Kapitza effect 
is found are relatively narrow with the widest being the first one from zero flux. Numerical investigations suggest that these intervals for the Kapitza effect shift to smaller values of fluxes, and the width of such intervals can even enlarge (see~\cref{fig: Phase diagram1}). Unlike the case of the well-known mechanical oscillator, within the time-scales separation approach, the effect can be observed for driving frequencies that are on the order of the natural frequency, $\omega \sim \omega_0$.

It is known that inverted regime of the oscillations of a dc SQUID, in which the global minimum shifts by $\pi$, is achieved by biasing the SQUID with proper amount of DC flux. Here we have shown that analogical regime can be achieved with purely AC magnetic flux. 

Our estimations suggest that the frequency of a LG beam should be in the microwave regime for the employed model to be correct. However, the effect might survive beyond the applicability of the model used. While the power output of current masers is much lower compared to that of lasers, we estimate that it is enough to produce the Kapitza effect. On the other hand, the model of the dynamics does not rely on the particular source of the AC magnetic flux, and the predicted dynamics can be tested in the lab with conventional electronics. The possibility of utilizing the Kapitza and inverted regime of a dc SQUID in applications, e.g., sensing, is not fully clear to us at this point.

The appearance of the local minimum in the effective potential allows to speculate about the possibility of utilizing it for constructing a novel Floquet-type qubit~\cite{gandon2022} based on Josephson junctions.

Based on a simple idea similar to the one explained in~\cref{sec: Effective JJ}, we expect the possibility of stabilizing the effective phase at $\phi_+ \neq 0,\pi$ in an asymmetric dc SQUID. We will investigate such a possibility in a follow-up paper.

\section{Acknowledgments}
We are grateful to A. Dalal, G. Jotzu, I. Khaymovich, D. Kuzmanovski, I. Sochnikov, and P. Volkov for useful discussions. 
This work was supported by the European Research Council ERC
HERO-810451 synergy grant, KAW foundation 2019-0068 and the Swedish Research Council
(VR). Work at University of Connecticut was supported by the OVPR  Quantum CT. 

\appendix

\section{A dc SQUID shunted by an external capacitor}
\label{app: dc-SQUID_shunted}

\begin{figure}[!htbp]
    \includegraphics[width=0.8\columnwidth]{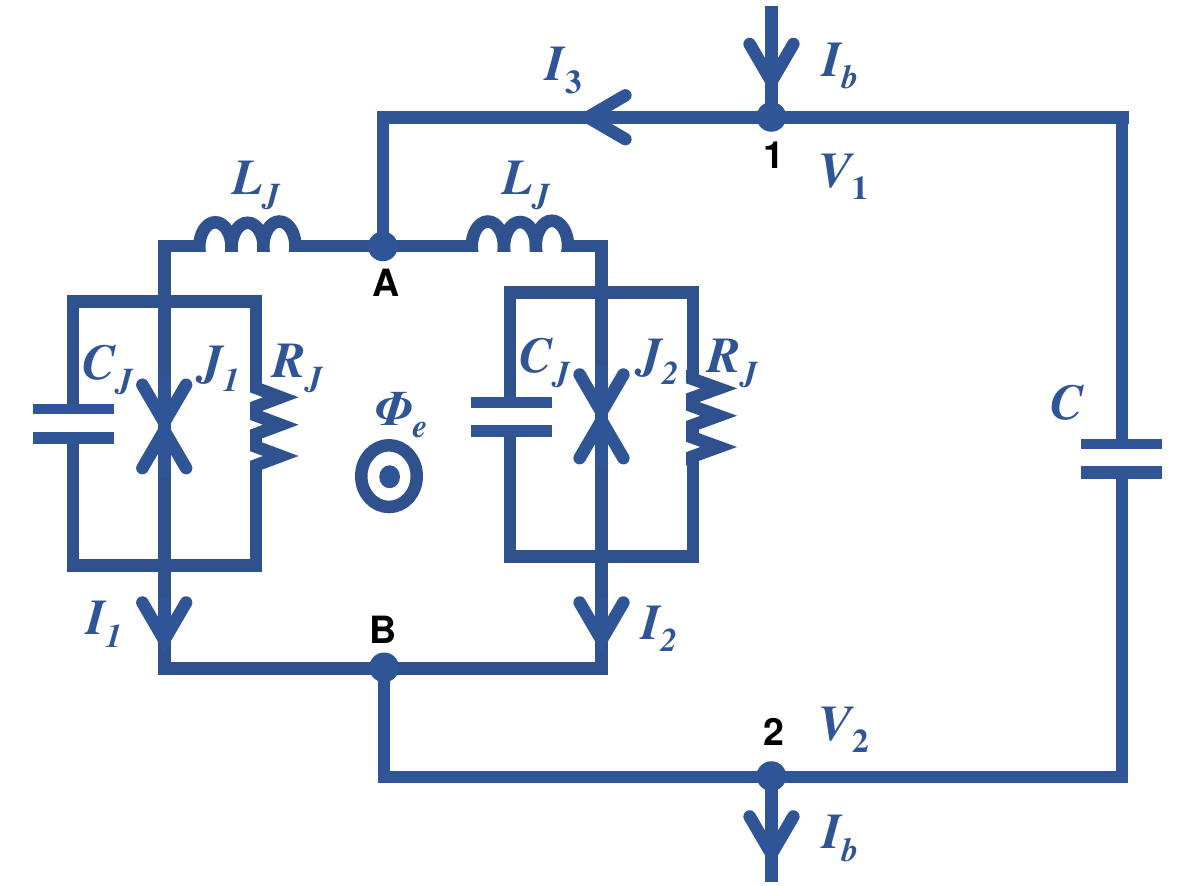}
    \caption{An equivalent scheme of a symmetric dc SQUID subjected to a LG beam and shunted by an external capacitor.}
    \label{fig: dc-SQUID shunted}
\end{figure}

In this section, we formulate the equations that govern the dynamics of a symmetric dc SQUID shunted by an external capacitor. The equivalent scheme is depicted in~\cref{fig: dc-SQUID shunted}. The flux equation is 
\begin{align}
    \phi_1-\phi_2 = \frac{2 \pi}{\Phi_0} \left( \Phi_e(t) - 2 L_J I_c \right),
\end{align}
where $I_c=\left( I_1 - I_2 \right)/2$ is the current circulating in the dc SQUID loop. The Kirchhoff's laws take the form
\begin{align}
\label{eq: B2}
    I_0 \sin \phi_1 + C_J \frac{\Phi_0}{2 \pi} \ddot \phi_1 + \frac{\Phi_0}{2 \pi R_J} \dot \phi_1 =& I_1, \\ 
\label{eq: B3}    
    I_0 \sin \phi_2 + C_J \frac{\Phi_0}{2 \pi} \ddot \phi_2 + \frac{\Phi_0}{2 \pi R_J} \dot \phi_2 =& I_2, \\
\label{eq: B4}    
    I_1 + I_2 =& I_3, \\
\label{eq: B5}
    I_3 + \dot Q =& I_b,
\end{align}
where $\phi_{1,2}$ are the Josephson phases of the JJs $J_{1,2}$, $I_{1,2}$ are currents through the JJs $J_{1,2}$, $I_3$ is the total current through the dc SQUID, $I_b$ is the bias current assumed to be constant, $I_0$, $C_J$, and $R_J$ are the critical current, capacitance, and the resistance of JJs, $L_J$ is the self-inductance of the parts with JJs of the circuit between points A and B (see~\cref{fig: dc-SQUID shunted}), and $Q = C \left( V_2 -V_1 \right)$ is the charge of the external capacitor $C$, where $V_{1,2}$ are the voltages at points $1,2$ (see~\cref{fig: dc-SQUID shunted}).

To find the expression for $V_1 - V_2$, let us consider two paths connecting points $1$ and $2$: $1-A-J_1-B-2$ and $1-A-J_2-B-2$. Then for $V_2$ we can write

\begin{align}
    V_2 = V_1 + \frac{\Phi_0}{2 \pi} \dot \phi_1 - L_J \dot I_1 - \frac{\dot \phi_e}{2}, \\ \nn
    V_2 = V_1 + \frac{\Phi_0}{2 \pi} \dot \phi_2 - L_J \dot I_2 + \frac{\dot \phi_e}{2}.
\end{align}
From which we find
\begin{align}
    V_2-V_1 = \frac{\Phi_0}{2 \pi} \frac{\dot \phi_1 + \dot \phi_2}{2} -\frac{L_J}{2} \dot I_3.
\end{align}
Using $\phi_{\pm} = \frac{\phi_1 \pm \phi_2}{2}$ as before, summing and taking the difference of~\cref{eq: B2,eq: B3}, and adding to them~\cref{eq: B5}, we get the following system of equations:

\begin{align}
\label{eq: B8}
    I_3 + C \frac{\Phi_0}{2 \pi} \ddot \phi_+ - \frac{C L_J}{2} \ddot I_3 = I_b, \\ 
\label{eq: B9}
    2I_0 \cos \pi \phi \sin \phi_+ + C_J \frac{\Phi_0}{\pi} \ddot \phi_+ + \frac{\Phi_0}{\pi R_J} \dot \phi_+ = I_3, \\
\label{eq: B10}
    2I_0 \cos \phi_+ \sin \pi \phi + C_J \Phi_0 \ddot \phi + \frac{\Phi_0}{R_J} \dot \phi = \frac{\Phi_0}{L_J} \left( \phi_e - \phi \right).
\end{align}

Substituting $I_3$ from~\cref{eq: B9} in~\cref{eq: B8}, we obtain a fourth order differential equation for $\phi_+$. This prohibits direct analysis of this system in terms of an effective potential. However, the idea of separation of time scales remains valid.

\bibliography{library}

\end{document}